# Ultralow 1/$f$ Noise in a Heterostructure of Superconducting Epitaxial Cobalt-Disilicide Thin Film on Silicon

*Shao-Pin Chiu,[†] Sheng-Shiuan Yeh,[†] Chien-Jyun Chiou,[‡] Yi-Chia Chou,[\*,‡] Juhn-Jong Lin,[\*,†,‡] and Chang-Chyi Tsuei[§]*

†Institute of Physics, National Chiao Tung University, Hsinchu 300, Taiwan

‡Department of Electrophysics, National Chiao Tung University, Hsinchu 300, Taiwan

§IBM Thomas J. Watson Research Center Yorktown Heights, NY 10598, U.S.A.

KEYWORDS: Superconducting silicide/silicon heterostructure, interfacial dynamic defects, 1/$f$ resistance noise, dimer reconstruction, two-level-system fluctuators

ABSTRACT: High-precision resistance noise measurements indicate that the epitaxial $CoSi_2$/Si heterostructures at 150 K and 2 K (slightly above its superconducting transition temperature $T_c$ of 1.54 K) exhibit an unusually low 1/$f$ noise level in the frequency range of 0.008–0.2 Hz. This corresponds to an upper limit of Hooge constant $\gamma \leq 3 \times 10^{-6}$, about 100 times lower than that of single-crystalline aluminum films on $SiO_2$ capped Si substrates. Supported by high-resolution cross-sectional transmission electron microscopy studies, our analysis reveals that the 1/$f$ noise is dominated by excess interfacial Si atoms and their dimer reconstruction induced fluctuators. Unbonded orbitals (*i.e.*, dangling bonds) on excess Si atoms are intrinsically rare at the epitaxial $CoSi_2$/Si(100) interface, giving limited trapping-detrapping centers for localized charges. With its excellent normal-state properties, $CoSi_2$ has been used in silicon-based integrated circuits for decades. The intrinsically low noise properties discovered in this work could be utilized for developing quiet qubits and scalable superconducting circuits for future quantum computing.

Superconducting circuit-based qubits technology represents the best hope of building a scalable fault-tolerant quantum computer.[1-3] The qubit coherence time has been increased dramatically from about 1 ns to 100 μs during the last decade.[3,4] Since quantum states are intrinsically fragile, errors are bound to occur in the course of any quantum computing process. Quantum error correction (QEC) is essential for quantum computation. However, even with the presently available superconducting qubits, tremendous complexity in developing a viable and feasible QEC architecture is unavoidable.[3,4] Even when the recently proposed surface code QEC approach is employed, one needs $10^3$ to $10^4$ physical qubits to define and operate one logical qubit.[5] This would involve a significant amount of extra interactions with the environment and represent sources of decoherence. To solve this daunting problem one must improve the coherence time at least by another factor of ten. To achieve this goal, except focusing on the improvement of clever qubit design and cryogenic microwave engineering, an alternative, valuable strategy is to find a practicable superconducting material which is essentially free of decoherence sources. In other words, it is desirable to fabricate and demonstrate a superconducting heterostructure with material properties superior to the commonly used $Al/AlO_x$-based heterostructures.

The coupling of a superconducting device with its environmental degrees of freedom results in detrimental fluctuations (noises) of charge, critical current or flux which, in turn, deteriorate the ultimate performance of the device. Recently, a variety of noise sources in superconducting quantum interference devices (SQUIDs) and qubits have been identified.[6,7] By and large, these decoherence sources manifest low-frequency noise in the form of a $1/f$ power spectrum density (PSD). Microscopically, they are believed to be closely associated with the defects residing at the superconductor/insulator interfaces and within the dielectric barrier.[7-9] Thus, an ideal candidate of such heterostructure should possess high-quality epitaxial interface with essentially no fluctuating atomic structures and charge trap centers. If charge transport in the epitaxial superconducting layer (in the normal state) is mainly limited by the boundary scattering at the superconductor/insulator interface,[10] the resistance noise measurements on such a heterostructure at $T > T_c$ can be used to characterize the dynamic properties of the interfacial structural defects as well as the low-frequency noise in superconducting devices such as qubits and SUQIDs.

The cobalt disilicide (CoSi$_2$), a compound with a sharp superconducting transition at $T_c \approx 1.5$ K,[11] is known for its excellent epitaxial properties on silicon,[12] single crystallinity with long carrier elastic mean free path,[10] robust covalent bond,[13] and relatively strong spin-orbit coupling.[14] Due to its good thermal stability and low resistivity ($\rho(300 \text{ K}) \approx 15 \text{ μΩ cm}$), it was widely used as interconnects and metallic contacts in CMOS and other electronic devices for years.[15] In this work, we show that this superconductor-silicon heterostructure possesses ultralow $1/f$ resistance noise at $T \geq 2$ K, and hence has high potential for applications in superconducting-circuit based quantum computing as well as other superconductive devices.

Since the 1970s the transition-metal disilicides, such as CoSi$_2$ and NiSi$_2$, have been widely used in the semiconductor (Si) circuit devices. Various synthesis methods and fabrication conditions were shown to sensitively influence the growth kinetics and microstructures of the disilicides formed on silicon.[12,15,16] Since CoSi$_2$ has a very small lattice mismatch with Si (−1.2%) and it is the only congruently melted intermetallic compound in the Co-Si alloy system, an epitaxial single-crystalline phase of $CoSi_2$ can be readily fabricated on a (100) or (111) Si substrate. In this study, we employ the standard deposition and thermal annealing technique to grow an epitaxial $CoSi_2$ film at the Co/Si(100) interface. We note that using the Si(100) substrates has twofold advantages. First, it is fully compatible with the present-day semiconductor integrated-circuit technology. Second, it leads to a dangling-bond-free $CoSi_2$ top surface,[17,18] and hence diminishing the $1/f$ noise sources in the $CoSi_2$/Si heterostructure. Consequently, the remaining possible noise sources are the defects at the epitaxial $CoSi_2$/Si(100) interface. This is the focus of the present work.

**RESULTS AND DISCUSSION**

Figure 1(a) shows the grazing x-ray diffraction (GIXRD) spectra for our samples, as indicated. The preferred orientations of (111), (220) and (311), manifesting good crystallinity, are clearly seen in epitaxial films. This result suggests that the $CoSi_2$ phase already formed with 30 min annealing at 800 °C. For comparison, Figure 1(a) also shows that such peaks are barely seen in a non-epitaxial $CoSi_2$ film nor a pure Co film. Transmission electron microscopy (TEM) studies indicate that the as-grown epitaxial $CoSi_2$ phase is single-crystalline along the direction normal to the film surface, while with a large grain size of ~ 300 nm in the lateral direction.[19] Figure 1(b) shows the variation of resistivity with temperature for the 30, 60 and 90 min annealed epitaxial $CoSi_2$ films, as indicated. The inset shows that the overall temperature dependences are similar, manifesting that of a good metal. The residual resistivities $\rho(2 \text{ K}) \approx 2.5 - 3 \text{ } \mu\Omega$ cm are low, while the $T_c$ is compatible to that of Al. In particular, a sharp superconducting transition occurs at 1.54 K in the 90 min annealed film. Such a high $T_c$ value can only be realized in $CoSi_2$ with the highest possible crystal quality.[20] In contrast, a two-step transition, accompanied with a transition tail, is found in the 30 (60) min annealed film, indicating that the sample structure is slightly inhomogeneous. Combining the resistivity results with high-resolution TEM studies (Figure 2), we can explain the two-step transition in terms of the presence of a Moiré fringe regime whose $T_c$ value is slightly reduced from 1.54 K. Figure 1(c) shows the superconducting transitions for a non-epitaxial and a 90 min annealed epitaxial CoSi$_2$ films, as indicated. A reduced $T_c$ value, accompanied by a broad two-step transition is evident in the non-epitaxial film, indicating significant inhomogeneous lattice structure.

Figure 1(d) shows the magnetoresistance (MR) for an epitaxial CoSi$_2$ film at 10 K. The magnetic field was applied perpendicular to the film plane. A positive, quadratic MR (see inset) was observed, suggesting a classical MR due to the Lorentz force and a long elastic mean free time/path of the charge carriers in the sample.[10,14] That is, the normalized MR is given by $[R(B) - R(0)]/R(0) \approx (\mu B)^2$, where $\mu = e\tau_e/m^*$ is the mobility, $e$ is the electronic charge, $\tau_e$ is the elastic mean free time, and $m^*$ is the effective carrier mass. In CoSi$_2$, $m^* \approx m_0$, the free electron mass.[14] The carrier (hole) concentration $n$ in CoSi$_2$ epitaxial films on Si(100) has been measured *via* Hall effect to be $n \approx 2.5 \times 10^{22}$ cm$^{-3}$,[14,21] giving rise to a Fermi energy of $E_F \approx 3.1$ eV and a Fermi velocity of $v_F \approx 1.05 \times 10^6$ m/s through the free-electron model. We then evaluated the elastic mean free path to be $l_e = v_F \tau_e \approx 120$ nm in the 90 min annealed epitaxial CoSi$_2$. From the film thickness dependence on residual resistivities and MRs (Supporting Information S1), we found a bulk value of the elastic mean free path being $l_e \approx 363$ nm in epitaxial CoSi$_2$/Si(100). Thus, surface/interface scattering plays a significant role in our 105 nm thick CoSi$_2$ films. On the other hand, we obtain a relatively short mean path of $l_e \approx 10$ nm in the non-epitaxial films. Table 1 lists the relevant parameters for various CoSi$_2$ films studied in this work.

Table 1. Relevant parameters for three epitaxial and one non-epitaxial CoSi$_2$ films.*

| Sample | $\rho$ (290 K) ($\mu\Omega$ cm) | $\rho$ (10 K) ($\mu\Omega$ cm) | $T_c$ (K) | $\Delta T_c$ (K) | $l_e$ (nm) | $t_M$ (nm) |
|---|---|---|---|---|---|---|
| 30 min annealed | 16.3 | 2.98 | 1.53 | 0.013 | 101 | 23 |
| 60 min annealed | 15.0 | 2.5 | 1.53 | 0.014 | 119 | 12 |
| 90 min annealed | 15.0 | 2.5 | 1.54 | 0.005 | 121 | 0 |
| non-epitaxial | 44 | 27.2 | 0.98 | 0.145 | 10 | – |

* $T_c$ is defined as the temperature corresponding to 50% of the normal-state resistance ($R_N$). The transition width $\Delta T_c$ is defined as the temperature difference between the $T$ values corresponding to 10% and 90% of $R_N$. $t_M$ is the thickness of Morie fringe region.

Figures 2(a)–(d) show the high-resolution cross-sectional transmission electron microscopy (XTEM) images of the CoSi$_2$/Si interface for the 30, 60, 90 and 90 min annealed films, respectively. It is clear that the CoSi$_2$ phase is epitaxial with the Si(100) substrate. On the CoSi$_2$ side, Moiré fringes are seen in the 30 and 60 min annealed films, but absent in the 90 min annealed films. The stripes in the right insets of Figures 2(a) and 2(b) show that the Moiré fringes extend for $\approx$ 23 and $\approx$ 12 nm from the interface in the 30 and 60 min annealed films, respectively. We ascribe the formation of the Moiré fringes to the lattice misfits and misorientations,[22] which induced strains near the CoSi$_2$/Si interface. In Figure 3, we present atomic simulations to illustrate that the Moiré fringes can be generated by rotating several CoSi$_2$ atomic layers against the bulk. In addition, in both 30 and 60 min annealed films, extra spots due to superlattice layers are seen in the diffraction patterns on the CoSi$_2$ side, see the left insets of Figures. 2(a) and 2(b), implying that the silicidation process was incomplete near the interface. With 90 min annealing, the phase transformation from Si to CoSi$_2$ was complete, and the strains were released by the formation of dislocations at the interface, rendering an atomically sharp interface without Moiré fringes, Figures. 2(c) and 2(d).

From the diffraction patterns shown in the insets of Figures 2(a)–(c), we found the expected epitaxy of $CoSi_2(100)$ on $Si(100)$ for all three annealing time periods, corresponding to the epitaxial relation $CoSi_2[011]$ // $Si[011]$ and $CoSi_2(200)$ // $Si(200)$. In Figure 2(d), another epitaxy of $CoSi_2(110)$ on $Si(100)$ was found, corresponding to the epitaxial relation $CoSi_2[011]$ // $Si[011]$ and $CoSi_2(022)$ // $Si(111)$.[12,15,22] The right inset of Figure 2(d) shows a zoom-in image for the latter epitaxial relation, highlighted with blue (Co) and yellow (Si) balls. These two kinds of epitaxial relation often compete in the $CoSi_2/Si(100)$ system. Judging from our fabrication conditions and the measured GIXRD results, we expect a small regime of our film possess the epitaxy of $CoSi_2(100)$ on $Si(100)$,[12,15,22] while the rest possess the epitaxy of $CoSi_2(110)$ on $Si(100)$. In the regime with the epitaxy of $CoSi_2(100)$ on $Si(100)$, a candidate unstable interfacial structure, the $2\times1$ dimer reconstruction, is known to exist,[23-27] which arises from the dimerization of excess Si at the interface.[23] Indeed, we found dimer reconstructions in the 90 min annealed sample, see the high magnification XTEM images in Figure 5(a) and further discussion below.

In the 30 min annealed film, pronounced Moiré fringes exist on the $CoSi_2$ side as indicated in the XTEM image of Figure 2(a). We replot Figure 2(a) into Figure 3(a) and show corresponding atomic simulations in Figure 3(b). The simulated fringes in Figure 3(b) reveal oblique stripes which were generated by a $50°$ rotational shift between two sets of $CoSi_2$ (011) planes with respect to the [011] axis. Furthermore, if structural defects such as edge dislocations are present in the $CoSi_2$ lattice, the generated Moiré fringes will contain distortions. This is indeed the case occurs in the dashed oval region in Figure 3(a). Corresponding atomic simulations are illustrated in the inset of Figure 3(b). In addition, atomic misfits at the $CoSi_2/Si$ interface can possibly cause kinked or stepwise boundaries, such as those indicated by the blue arrow in Figure 3(a) and correspondingly by the red arrows in Figure 3(b). From our electrical-transport studies and the parameters listed in Table 1 for the three epitaxial films, we clearly see that the presence of more pronounced Moiré fringes gives rise to higher residual resistivities, shorter elastic mean free paths, and wider superconducting transition widths.

The imperfection scattering centers in a metallic film can be categorized into two groups: static defects and dynamic defects (or dynamical structural defects). In the high-quality epitaxial CoSi₂/Si system, we expect that both types of defects are rare and reside mainly at the interface. Therefore, the charge transport properties are essentially governed by the boundary scattering, because, as aforementioned, the carrier elastic mean free path extracted from MRs is mainly limited by the film thickness. With this property the microscopic properties of interfacial dynamic defects can be detected through the measurement of resistance fluctuations.

We have carried out resistance noise studies of epitaxial CoSi₂ films using a high-resolution ac bridge technique schematically shown in Figure 4(a) (also see Methods). The dynamical structural defects can be modeled as two-level systems (TLSs) which induce fluctuations of scattering cross sections owing to their repeated relaxation processes.[28,29] It is well known that if the distribution of the TLS activation energy is broad and flat, the voltage noise power spectrum density (PSD), $S_V$, for an ohmic conductor carrying a constant current would show $1/f$ dependence. This is widely described by the empirical Hooge expression[28,29]

$$S_V = \gamma \frac{V^2}{N_c f} + S_V^0 \qquad (1)$$

where the first term is the voltage noise PSD of the sample, and the second term is the background noise including the Johnson-Nyquist noise and the electronic noise from the measurement circuit. $\gamma$ is the dimensionless Hooge parameter, $V$ is twice of the root-mean-square voltage drop $\langle V_1 \rangle$ across one half of the sample (Figure 4(a)), and $N_c$ is the total number of charge carriers in the sample.

In the following discussion, we focus on three types of 90 min annealed samples: sample type A (C) is an epitaxial CoSi$_2$ film without (with) dilute HF treatment of the CoSi$_2$ top surface before depositing Ti/Au electrodes, and sample type B is a non-epitaxial CoSi$_2$ film with dilute HF treatment. The dilute HF treatment was aimed at removing the thin native oxide layer on the CoSi$_2$ surface in order to ensure a good electrical contact between CoSi$_2$ and Ti/Au. The measured voltage noise PSDs of samples type A and type C at 150 K are shown in Figures. 4(b) and 4(d), respectively. The background PSD (black curves) $S_V(V_1 = 0) \approx 6 \times 10^{-18}$ V$^2$/Hz indicates the resolution limit of our setup, which is just the input noise level of the preamplifier. In Figures. 4(b) and 4(d), the PSDs at finite bias voltages reveal $1/f$ dependence in the frequency range of $\sim 0.06 - 2$ Hz and $\sim 0.008 - 0.2$ Hz, respectively. For both types of samples, we plot the variation of $\langle fS_V \rangle$ with $V^2$ in Figures. 4(c) and 4(e). Linear dependence of $\langle fS_V \rangle$ on $V^2$ is evident, as expected from the prediction of Eq. (1). However, it should be emphasized that the fitted slopes, $i.e.$, the $\gamma / N_c$ values, for samples type A and type C are distinctly different. We obtain $\gamma \approx (1.4 \pm 0.2) \times 10^{-2}$ for sample type A, as compared to the extremely small value of $\gamma \approx (2.8 \pm 0.3) \times 10^{-6}$ for sample type C.

In summary, Figure 4(f) shows the normalized voltage noise PSD, $N_c S_V / V^2$, for three types of samples measured at 150 K, as indicated. Also included are representative data for a polycrystalline Al film at 300 K (taken from ref. 32) as well as a single-crystalline Al film at 150 K (taken from ref. 33). In all cases, the $f^{-1}$ frequency dependence (dashed and dotted lines) is evident. The measured noise levels of these samples differ significantly and are sensitive to sample preparation conditions. The larger PSD signal found in sample type A resulted from the large contact noise due to the highly resistive native oxide layer. The extracted $\gamma$ value is $\sim 10^{-3} - 10^{-2}$. For sample type B, a non-epitaxial CoSi$_2$ film with dilute HF treatment, a notable decrease by $\sim 2$ orders of magnitude in the noise PSD was found ($\gamma \sim 1 \times 10^{-4}$). For sample type C with dilute HF treatment, the measured noise PSD is even lowered by another two orders of magnitude ($\gamma \approx 3 \times 10^{-6}$). Such an extremely low level of noise definitely deserves attention. First, we point out that our sample resistance is small ($R < 100\ \Omega$), rendering a very low thermal noise PSD ($S_{V\_thermal} = 4 k_B T R \leq 8.3 \times 10^{-19}\ \text{V}^2/\text{Hz}$ at 150 K, where $k_B$ is the Boltzmann constant). Thus, the flattening in the PSD of curve C at $f \geq 1$ Hz actually originated from the noise level of our measurement electronics (*i.e.*, the preamplifier) rather than from the sample itself. In other words, due to the noise PSD of the preamplifier ($S_V (V_1 = 0) \sim 6 \times 10^{-18}\ \text{V}^2/\text{Hz}$), this extracted $\gamma$ value is really an upper bound rather than an exact value. As expected, our measurements carried out at 2 K (a temperature just above $T_c$) also indicated that the Hooge constant $\gamma \leq 3 \times 10^{-6}$. Although a low $\gamma$ value of $\sim 10^{-6}$ was found in semiconductors, such as high-quality GaAs/AlGaAs heterostructures[30] and well-annealed doped silicon,[31] it has not been found in metals or superconductors (in the normal state) except the present epitaxial CoSi$_2$/Si(100) heterostructures. We should emphasize that this upper bound $\gamma$ value is nearly two (three) orders of magnitudes lower than that of single-crystalline[32] (polycrystalline[33]) Al films, and eight orders of magnitude lower than that ($\approx 7.7 \times 10^2$) in the high-temperature superconductor YBa$_2$Cu$_3$O$_{7-\delta}$ single crystals.[34] We mention that this measured value of $\gamma \approx 3 \times 10^{-6}$ is about two orders of magnitude higher than the theoretical estimate of the quantum $1/f$ noise.[35] Recall that polycrystalline Al films are an indispensable material widely used in present-day SQUIDs and qubits. Thus, our finding of the ultralow $1/f$ noise in epitaxial CoSi$_2$/Si(100) heterostructure with $T_c \approx 1.5$ K may likely constitute a feasible substitute for current superconducting devices. Furthermore, it should be reiterated that our choice for using the Si(100) substrates is readily compatible with existing semiconductor integrated-circuit technology.

In the epitaxial CoSi$_2$/Si(100) heterostructures, the possible origins of the TLSs are unstable interfacial structures and fluctuations of the number of localized charges at the interface/surface. Combining high-resolution XTEM studies, we find that the responsible TLSs are dimer reconstructions and dangling bonds of excess Si atoms residing at the interface (see further discussion below). The amount of these interfacial TLSs governs the total fluctuating scattering cross sections and determines the Hooge $\gamma$ value. Therefore, we may estimate the TLS density, $n_{\text{TLS}}$, as follows.

The relation between the Hooge parameter $\gamma$ and the TLS density $n_{\text{TLS}}$ is given by[36]

$$\gamma = \frac{\left( n l_e \beta \sigma \right)^2 \left( n_{\text{TLS}} / n \right)}{\ln \left( \omega_{\max} / \omega_{\min} \right)} \qquad (2)$$

where $\beta = \delta\sigma / \langle \sigma \rangle$, $\delta\sigma$ denotes the fluctuating scattering cross section, and $\langle \sigma \rangle$ denotes the averaged scattering cross section. $\omega_{\max}$ and $\omega_{\min}$ are the upper- and lower-limit frequencies, respectively, of the TLSs. We can rewrite the above equation into the following form

$$\gamma = \frac{\beta^2 / e^4}{\ln \left( \omega_{\max} / \omega_{\min} \right)} \left( n_{\text{TLS}} \frac{(m^*)^2}{n} \frac{v_F}{\rho^2} \sigma^2 \right) \qquad (3)$$

Typically, $\beta \approx 0.2$ in real materials.[36] The value of $\ln \left( \omega_{\max} / \omega_{\min} \right)$ can be taken to be approximately the same for all materials due to its logarithmic dependence. Thus, from the known experimental values of $\gamma \approx 1 \times 10^{-3}$ in polycrystalline aluminum[32,33] and $n_{\text{TLS}} \approx 10^{22} - 10^{24}$ m$^{-3}$ in typical polycrystalline metals and alloys,[37-40] we have evaluated a very low TLS density of $n_{TLS} \approx 10^{18} - 10^{20}$ m$^{-3}$ in our epitaxial CoSi$_2$ films, using the measured upper-bound value of $\gamma \approx 3 \times 10^{-6}$.

We obtain a very small value of $n_{TLS} \sim 10^{18} - 10^{20}$ m$^{-3}$, which is $\sim$ 10,000 times lower than that in polycrystalline Al films. In Table 2, we compare the most relevant parameters for epitaxial CoSi$_2$/Si(100) heterostructures and single-crystalline and polycrystalline Al films grown on SiO$_2$ or sapphire.[32,33] The advantages of epitaxial CoSi$_2$ over Al are evident. An ultra-low level of $n_{TLS}$ immediately suggests the high potential of applying epitaxial CoSi$_2$ films in high-Q-factor superconducting resonators and superconducting qubits which demand long coherence time. As mentioned, since the charge carriers in our films have a long bulk mean free path, they essentially undergo elastic scattering only at the interface/surface. Therefore, the TLSs in our films, if exist, must most likely be associated with interfacial defects. Then, we may calculate the areal TLS density from the bulk density, $i.e.$, $n_{TLS\_2D} = n_{TLS} \times$ (film thickness) $\approx 10^{11} - 10^{13}$ m$^{-2}$. In the case of single crystalline Al films,[32] the resistance noise most likely arises from the TLSs residing at the Al/SiO$_2$ and Al/AlO$_x$ interfaces (a surface AlO$_x$ layer naturally formed in atmosphere), where the value of $n_{TLS\_2D}$ extracted from Hooge $\gamma$ is $\sim 3 \times 10^{14} - 3 \times 10^{16}$ m$^{-2}$. On the other hand, the $n_{TLS\_2D}$ value deduced from flux noise measurements on Al/AlO$_x$-based SQUIDs is $\sim 10^{17}$ m$^{-2}$,[8,9] in satisfactory consistency with the value of $n_{TLS\_2D}$ from $\gamma$. Thus, we can reasonably estimate that, if a SQUID structure is made of CoSi$_2$/Si and the flux noise measured, the magnitude of $n_{TLS\_2D}$ to be deduced would be on the order of $\sim 10^{11} - 10^{13}$ m$^{-2}$, according to the Hooge $\gamma$ value given above. Note that such an areal density is 4 to 6 orders of magnitude lower than that ($\sim 10^{17}$ m$^{-2}$) in the present-day SQUIDs and qubits made of Al/AlO$_x$.[8,9] For the Si(100) plane, the Si density is

$\sim 7 \times 10^{18}$ m$^{-2}$. Our $n_{TLS\_2D}$ value thus suggests that only an extremely small fraction of $\leq 1$ ppm of the interfacial Si atoms act as TLSs. In the rest of the paper, we elaborate on the TLS picture to analyze and explain in detail the origin for the ultralow noise in epitaxial CoSi$_2$/Si(100) heterostructures.

Table 2. Parameters of epitaxial CoSi$_2$ and Al.*

| Material | $\rho$(300 K) ($\mu\Omega$ cm) | $T_c$ (K) | Hooge $\gamma$ | $n_{TLS}$ (m$^{-3}$) |
|---|---|---|---|---|
| Epitaxial CoSi$_2$ | 15 | 1.54 | $\leq 3 \times 10^{-6}$ | $\leq 10^{18} - 10^{20}$ |
| Single-crystalline Al | 2.8 | $\sim 1.2$ | $1 \times 10^{-4}$ | $10^{21} - 10^{23}$ |
| Polycrystalline Al | 2.9 | $\sim 1.2$ | $1 \times 10^{-3}$ | $10^{22} - 10^{24}$ |

*The Hooge $\gamma$ values for single-crystalline and polycrystalline Al are taken from refs. 32 and 33, respectively.

At the CoSi$_2$(100)/Si(100) interface, a well-known unstable structure is the $2 \times 1$ dimer reconstruction, as previously established by the specific XTEM studies.[23-25] In Figure 5(a), the (1) arrows point to the regimes where the interfacial Co atoms are sixfold coordinated with Si by robust covalent bonds.[13] In this case, the atomic structure is most stable, as schematically depicted by the model in Figure 5(b). On the other hand, the (2) arrow points to a bright atomic cluster where two excess Si atoms can exist. As a consequence, the excess Si atoms can bond to form a dimer which would lower the interfacial energy and reduce the number of dangling bonds (DBs) by half (DBs are schematically depicted by short red dotted lines in Figure 5.) Indeed, this kind of dimer reconstructions at the CoSi$_2$/Si interface was first found and addressed by Loretto *et al*.[23] (see further discussion on boundary lattice structure in Supporting Information S2). In Figure 5(c) we present a 3D model for the interfacial structure with a dimer (depicted by a red bar in Figure 5). The medium-energy ion scattering studies of the CoSi$_2$/Si(100) interface[24] have revealed that an asymmetric dimer structure, Figures. 5(f) or 5(g), is energetically more favorable than a symmetric structure, the lower panel of Figure 5(d). This situation is reminiscent to that of the well-known buckled dimers on Si(100) surfaces.[41-43] The two asymmetric dimer configurations represented in Figures. 5(f) and 5(g) are degenerate and called the $\theta_0$ and $-\theta_0$ states, respectively. In terms of the TLS model, these two states correspond to the two minima of the double-well potential shown in Figure 5(e). Here the $\theta = 0$ state stands for a symmetric dimer, with a potential barrier height $V_B$. At low temperatures (*e.g.*, close to $T_c$), the dimer structure will quantum-mechanically switches back and forth between the $\theta_0$ and $-\theta_0$ states,[41-46] generating an important characteristic energy scale, $\Delta_0$, between the ground state and the first excited state.

Now we show that the TLS model of Figure 5(e) can well explain the ultralow $1/f$ noise found in this work. Microscopically, when the charge carriers undergo boundary scatterings, where the TLSs reside, the $1/f$ noise is generated. The characteristic switching rate of the buckled dimers at the CoSi$_2$/Si(100) interface can be evaluated through the well-known relation[29]

$$\Delta_0 / h = f_0 \cdot \exp(-2\pi d \sqrt{2mV_B} / h) \qquad (4)$$

where $f_0$ is the typical vibrational frequency, $m$ is the mass of the switching object, $d$ is the distance between the two potential minima, and $h$ is the Plank constant. At the CoSi$_2$/Si interface, $f_0 \sim 1.3 \times 10^{13}$ Hz is the Debye frequency of silicon, $m$ is the mass of Si atom, $d$ is the displacement of flipping Si atoms ($\sim 0.6$ Å for $\theta_0 \approx 15°$ and bond length $\approx 2.3$ Å),[23,42] and $V_B \sim 0.1 - 0.26$ eV for the hole doping circumstance.[46] With these values, we readily obtain $\Delta_0 / h = 5 \times 10^{-3} - 4 \times 10^3$ Hz. Note that this range of switching rate exactly covers the frequency range for the $1/f$ noise data shown in Figure 4(f). That is, the dimer reconstruction due to excess

Si at the CoSi$_2$/Si(100) interface is the responsible TLSs. They cause the measured 1/$f$ noise. We mention that a similar switching rate of the buckled dimers on Si(100) surface has recently been inferred from the flicker noise in the tunneling current between a scanning tunneling microscopy (STM) tip and a fluctuating dimer.[44] Apart from the fluctuating dimers discussed above, the remaining dangling bonds (DBs) associated with the excess Si atoms may form another type of TLSs, because the $sp^3$ orbitals of Si DBs can act as trapping-detrapping centers for localized charges.[47,48] However, since in this case the tunneling object is an electron, the typical trapping-detrapping rate ($\Delta_0/h \sim 10^{13}$ Hz )[48] is too fast to be captured by 1/$f$ noise measurements, and thus this process can be ruled out in our case. Thus, we conclude that the fluctuating dimers are the responsible mechanism for generating the 1/$f$ noise in epitaxial CoSi$_2$/Si(100) heterostructures. The magnitude of the noise is ultralow because we have an extremely low areal (dimer) density of $n_{TLS\_2D} = 10^{11} - 10^{13}$ m$^{-2}$. The areal density of the remaining DBs must be also on the same order, because there are two remaining DBs per dimer. For comparison, we should note that, in present-day Al/AlO$_x$–based superconducting qubit devices, large amounts of trapped charges in DBs in the dielectric regime often generate notoriously high levels of 1/$f$ noise.[7-9]

**CONCLUSION**

In summary, we have found ultralow 1/$f$ noise in a heterostructure of superconducting epitaxial CoSi$_2$ thin film on Si(100). The noise level is 2 to 3 orders of magnitude lower than that in Al, while a sharp superconducting transition occurs at a manageably high temperature of 1.5 K. The microscopic origin for the ultralow 1/$f$ noise is ascribed to the dimer reconstruction at the CoSi$_2$/Si(100) interface. The areal density of fluctuating dimers was evaluated and found to be 4 to 6 orders of magnitude lower than those in the widespread superconducting qubit devices made of Al/AlO$_x$. While preparing an epitaxial dielectric barrier in Al/Al$_2$O$_3$–based structures to reduce the $1/f$ noise level is nontrivial,[6,49,50] it is highly workable to utilize single-crystalline, undoped Si as an epitaxial dielectric layer, sandwiching it between two $CoSi_2$ films to form a (lateral) Josephson junction or SQUID with a one-step lithographic process. One can then study the flux noise spectra of a $CoSi_2$/Si-based SQUID to justify the low noise property. In addition, measurements of the relaxation time $T_1$ should be performed to check if any source(s) for high-frequency relaxation are rare. Another study can be done is to make a superconducting coplanar waveguide using epitaxial $CoSi_2$/Si to check whether it has a high Q-factor. Although the $1/f$ noise issues can be minimized in certain qubit designs,[51,52] the noisy superconducting components still exist in other parts of superconducting devices and play an important role in determining the performance of the whole devices/circuits.

It should be stressed that the $1/f$ noise level will hinder the performance of the ultra-sensitive magnetic field sensors based on SQUIDs, where the sensitivity cannot be improved by increasing the measurement time.[53,54] Thus, a very low $1/f$ noise level for the superconducting materials and qubits is definitely required. At present, the bottleneck for superconducting circuit-based quantum computing lies at the fourth stage, where the critical difficulty is how to maintain sufficiently long lifetime for logical quantum memory.[3] We propose to use the excellent epitaxial properties of $CoSi_2$/Si to minimize decoherence sources at the surfaces and interfaces, and hence increase the coherence time for breaking through the fourth stage. Note that the fabrication processes for $CoSi_2$/Si devices are readily compatible with the present-day silicon-based technology. These new-generation $CoSi_2$/Si-based devices may be exploited as potential substitutes for the current $Al/AlO_x$-based devices.

**METHODS**

**Sample Preparation.** Epitaxial $CoSi_2$/Si(100) heterostructures. A 30 nm thick Co film was first deposited *via* thermal evaporation on a native oxide removed lightly boron-doped (*p*-type) Si(100) substrate. The Co/Si specimen was subsequently annealed at 800ºC in a high vacuum of $1 \times 10^{-5}$ torr for 30, 60 or 90 min. Such extended annealing time periods, as compared with the rapid thermal processes (RTP) commonly used in the CMOS production,[12,15,55] ensured complete reaction of the deposited cobalt atoms with substrate Si atoms. The thickness of the $CoSi_2$ film thus formed was $\approx 105$ nm, in good agreement with previous finding.[55] By employing three periods of annealing time, we were able to examine various stages of the phase formation and morphology at the $CoSi_2$/Si interface.

Non-epitaxial $CoSi_2$ films. For comparison, we also grew non-epitaxial $CoSi_2$ films and studied their structural, electrical and noise properties. A Co/Si/Co/Si/Co/Si multilayer was first deposited on a 300-nm $SiO_2$ capped Si substrate *via* thermal evaporation, where each Co (Si) layer was 10 (36) nm thick. This thickness ratio of 1:3.6 facilitated the formation of a stoichiometric $CoSi_2$ film when the multilayer was annealed in a high vacuum at 800ºC for 90 min. The as-grown films were polycrystalline, with a gran size of $\approx 10$ nm. The $CoSi_2$/$SiO_2$ interface did not reveal any epitaxial relations, as evidenced from XTEM studies.

**High-Resolution Cross-Sectional Transmission Electron Microscopy (XTEM) Studies.** The XTEM studies of our films were carried out using a spherical aberration corrected scanning transmission electron microscope (JEOL ARM200F).

**Electrical-Transport Measurements.** The resistivity $\rho(T)$ of our $CoSi_2$ films was measured using an ac resistance bridge (Linear Research LR700 or LR400) with the four-probe configuration in an Oxford Heliox $^3$He cryostat. The $T_c$ value was determined by the 50% drop of the normal-state resistance, $R_N$. The superconducting transition width $\Delta T_c$ was defined by the temperature difference between the 10% and 90% drops of $R_N$. The $T_c$ and $\Delta T_c$ values are listed in Table 1.

**Low-Frequency Noise Measurements.** The $1/f$ noise of our films was measured with an ac bridge technique using the five-terminal configuration,[56,57] Figure 4(a). The CoSi$_2$ films were patterned by the electron-beam lithography technique to have typical dimensions of $\approx 50$ μm long and $\approx 0.8$ μm wide. The electrodes were made of Ti/Au (15/65 nm) *via* thermal evaporation. Two current leads ($I_1$ and $I_2$) were respectively connected to two variable resistors ($R_1$ and $R_2$), while two voltage leads ($V_1$ and $V_2$) were connected to the input of a preamplifier operating in the differential mode. The resistances for the two CoSi$_2$ segments were approximately equal ($r_{s1} \approx r_{s2}$). The two variable resistors were tuned to balance the two voltages $V_1$ and $V_2$. When the bridge was in balance, the average voltage drop $\langle \delta V \rangle \approx \langle V_1 - V_2 \rangle$ was minimized so that the higher gain of the preamplifier could be utilized to amplify the signal against the contingent noise in the following stages. The balance of the bridge also eliminated the noise contribution from the voltage fluctuations of the power supply and the contact resistance in the central terminal. In our measurements, the unbalance ratio was $\langle \delta V \rangle / \langle V_1 \rangle \approx 4 \times 10^{-4} - 1 \times 10^{-2}$ for all samples. A sinusoidal driving voltage was used to modulate the low frequency signal of our samples to an optimal frequency ($\approx 3.1$ kHz) of the preamplifier (Stanford SR560), where the $1/f$ noise from the preamplifier was minimized. The amplified signal was thus demodulated using the phase-sensitive-detecting technique to retrieve the $1/f$ signal of the sample. The retrieved signal was digitalized by a lock-in amplifier (Stanford SR830) with a rate of 512 Hz and recorded by a spectrum analyzer (Stanford SR785). Throughout measurements, we set $R_i > 10 r_{si}$ ($i = 1$, 2). Typical sets of raw data are shown in Figures. 4(b) and 4(d).

ASSOCIATED CONTENT

**Supporting Information:** Superconducting transition and magnetoresistance for $CoSi_2/Si(100)$ heterostructures; Assignment of atomic positions and atomic model for the $CoSi_2/Si(100)$ interface. This material is available free of charge *via* the Internet at http://pubs.acs.org.

AUTHOR INFORMATION

**Corresponding Author**

*Email: ycchou@nctu.edu.tw
*Email: jjlin@mail.nctu.edu.tw

**Author Contributions**

S.P.C. synthesized the samples and carried out transport characterizations, S.S.Y. performed $1/f$ noise measurements, C.J.C. and Y.C.C. carried out TEM studies, J.J.L. and C.C.T. supervised the work. All authors participate in the discussion and interpretation of the results and co-write the manuscript.

**Notes**

The authors declare no competing financial interests.

ACKNOWLEDGMENT

The authors thank Ruey-Tay Wang for experimental assistance of $1/f$ noise measurements. This work was supported by the Taiwan Ministry of Science and Technology (MOST) through Grant numbers MOST-103-2112-M-009-017-MY3, 105-2923-M-009-005-MY2 (J.J.L.), and MOST-104-2112-M-009-015-MY3 (Y.C.C), and the Taiwan Ministry of Education (MOE) ATU Plan.

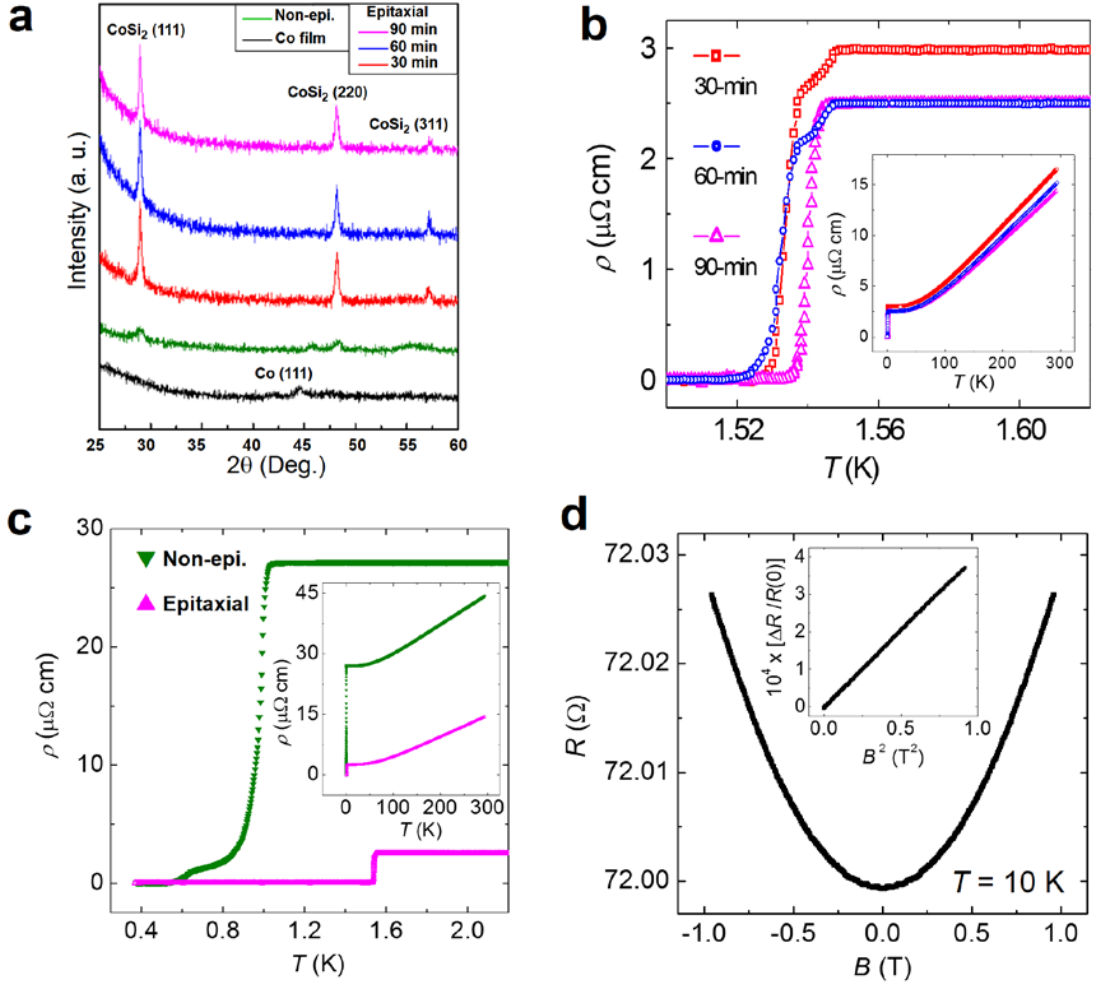

**Figure 1.** Grazing x-ray diffraction (GIXRD) spectra, cross-sectional TEM (XTEM) images, superconducting transitions, and magnetoresistance (MR) of $CoSi_2$ films. (a) GIXRD spectra for 30, 60 and 90 min annealed epitaxial $CoSi_2$ films, as indicated. The film thickness is ≈ 105 nm. The spectra for a non-epitaxial $CoSi_2$ and a pure Co films are also shown for comparison. (b) Superconducting transitions for three epitaxial $CoSi_2$ films, as indicated. The 90 min annealed film undergoes a very sharp superconducting transition at 1.54 K, with a narrow transition width of 5 mK. The inset shows the variation of resistivity with temperature between 1 and 300 K. (c) Superconducting transitions for a non-epitaxial and an epitaxial $CoSi_2$ films, as indicated. Inset: temperature dependence of resistivity between 1 and 300 K for the two films. (d) MR for a 90 min annealed epitaxial $CoSi_2$ film at $T$ = 10 K. The inset shows that the normalized MR is proportional to the square of magnetic field.

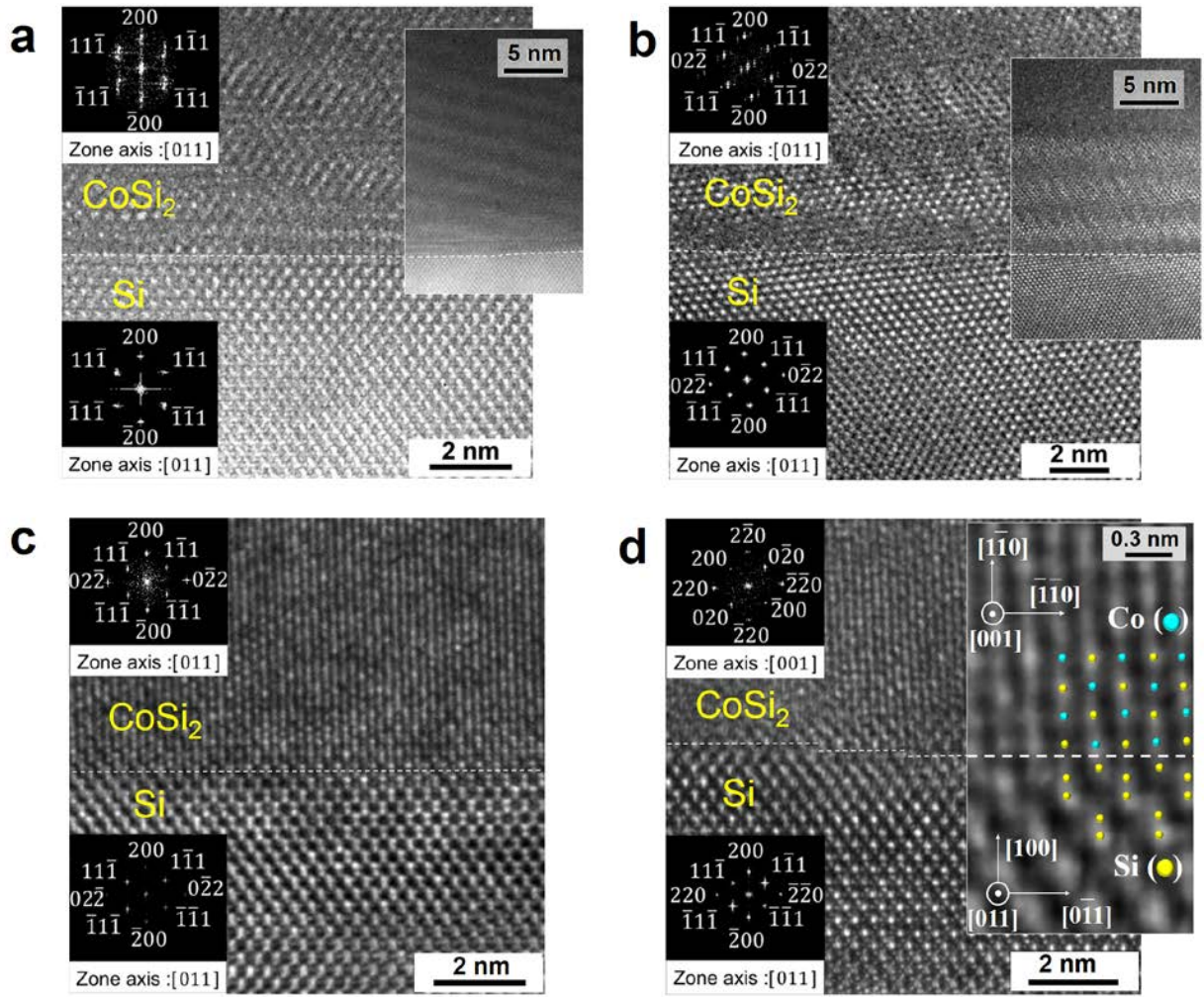

**Figure 2.** (a)–(d) show the XTEM images for 30, 60, 90 and 90 min annealed epitaxial CoSi$_2$ films, respectively. In each figure, the dashed line indicates the CoSi$_2$ and Si phase boundary, and the left insets show the diffraction patterns for (top) CoSi$_2$ and (bottom) Si(100). The right insets in (a) and (b) show XTEM images for Moiré fringes. The right inset in (d) shows a zoom-in image for the epitaxy of CoSi$_2$(110) on Si(100), highlighted with blue (Co) and yellow (Si) balls.

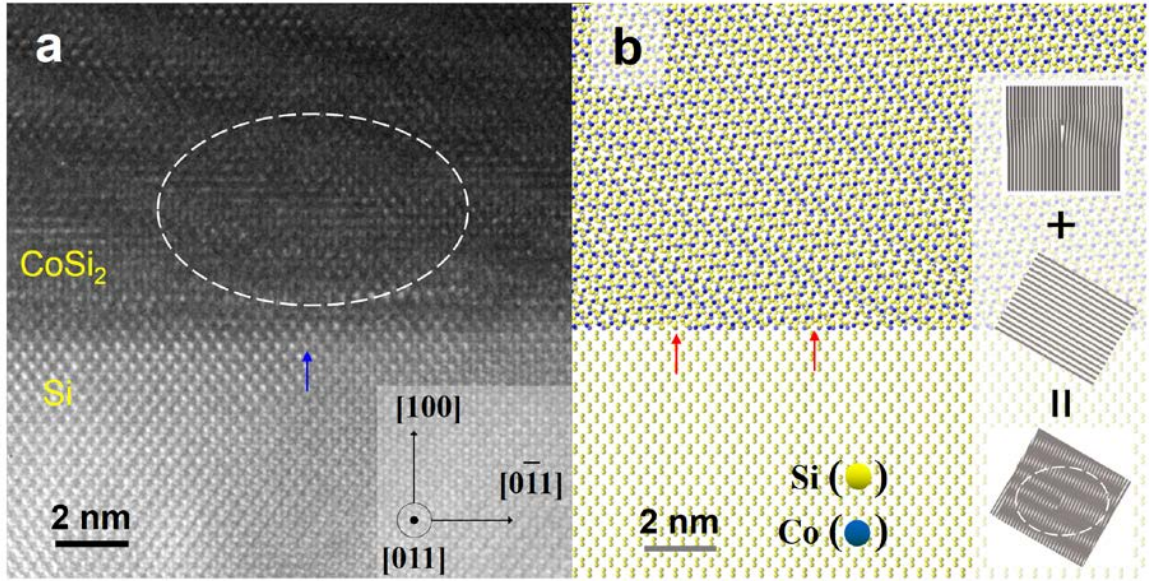

**Figure 3.** (a) XTEM image of a 30 min annealed sample. Moiré fridges on the CoSi₂ side are evident. (b) Corresponding atomic simulations for the image of Moiré fringes shown in (a). The inset shows a schematic for a distorted Moiré fringe due to the existence of an edge dislocation combined with a rotational shift between two sets of CoSi₂ lattice. The arrows in (a) and (b) indicate three interfacial locations with notable atomic misfits at the CoSi₂-Si boundary.

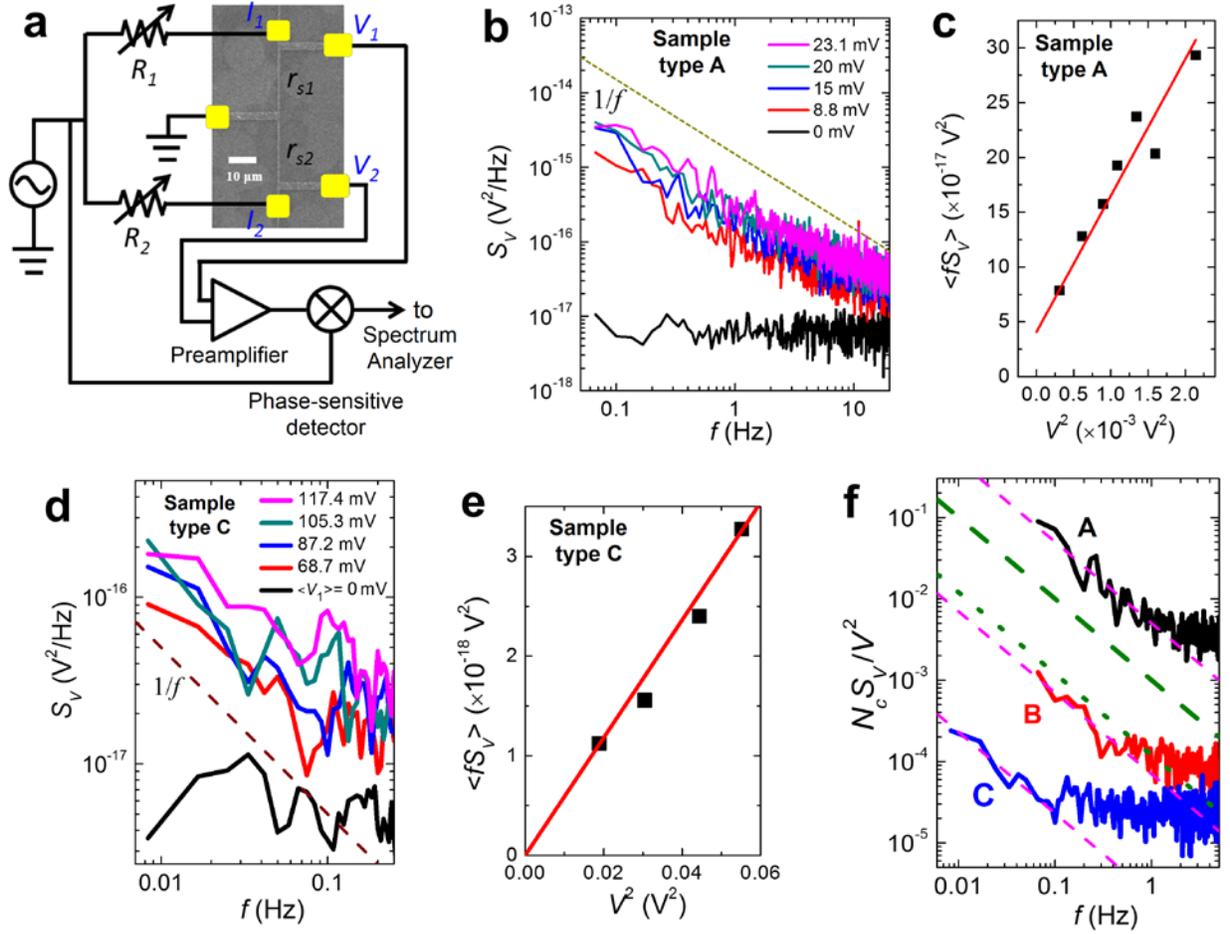

**Figure 4.** (a) A schematic for $1/f$ noise measurement setup. The five-terminal CoSi$_2$ sample (0.8 μm wide and ≈ 105 nm thick) was fabricated *via* electron-beam lithography. The yellow regions schematically depict Ti/Au patterns on CoSi$_2$ as electrodes and wire-bonding pads. (b) and (d) are the voltage power spectrum densities (PSD) of samples type A and type C, respectively, under various bias voltages as indicated. The dashed lines are drawn proportional to $1/f$ and is a guide for the eye. The $\langle fS_V \rangle$ values are calculated from the $1/f$ dependent regime and plotted as a function of $V^2$ for samples type A and type C in (c) and (e), respectively. The straight lines are linear fits. (f) Normalized voltage noise PSD for three types of samples measured at 150 K: sample type A (C) is an epitaxial CoSi$_2$ film without (with) HF treatment of the CoSi$_2$ top surface before depositing Ti/Au electrodes, and sample type B is a non-epitaxial CoSi$_2$ film with HF treatment. The normalized noise PSDs for a polycrystalline Al film at 300 K (green dashed line) (taken from ref. 32), and a single-crystalline Al film at 150 K (green dotted line) (taken from ref. 33) are plotted

for comparison. The dashed and dotted lines indicate $1/f$ frequency dependence. The data of sample type C suggests an ultralow Hooge parameter $\gamma \leq 3 \times 10^{-6}$ for the epitaxial $CoSi_2/Si(100)$ heterostructure.

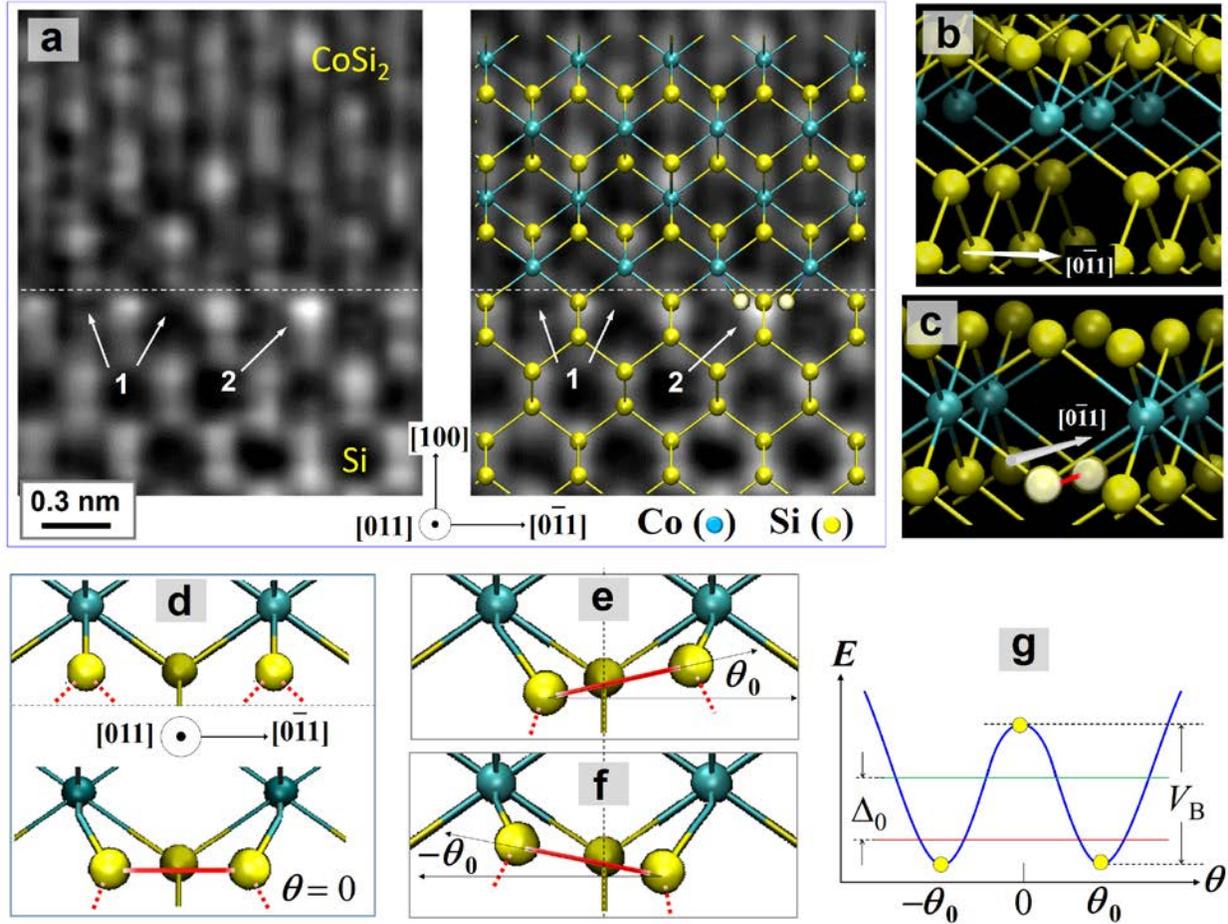

**Figure 5.** XTEM images and schematic diagrams depicting two-level system models for fluctuating dimers. (a) Left panel: XTEM image with epitaxy of $CoSi_2(100)$ on $Si(100)$ taken from a portion of Figure 2(c). Right panel: the same image with the mapping of a ball-and-stick model projected on the (011) plane. The region enclosed by (1) arrows can be modeled with sixfold coordinated Co, as shown in (b). The regions enclosed by (2) arrows can be modeled with a dimer reconstruction (red bar), as shown in (c). The upper panel in (d) depicts a model without a dimer reconstruction. The lower panel of (d), (f) and (g) depict symmetric ($\theta_0 = 0$) and buckled ($\theta_0, -\theta_0$) dimers, respectively. In the double-well potential shown in (e), the two minima represent the buckled dimer states, and the top of barrier height $V_B$ represents the symmetric dimer state.

**Table of Contents Graphic:**

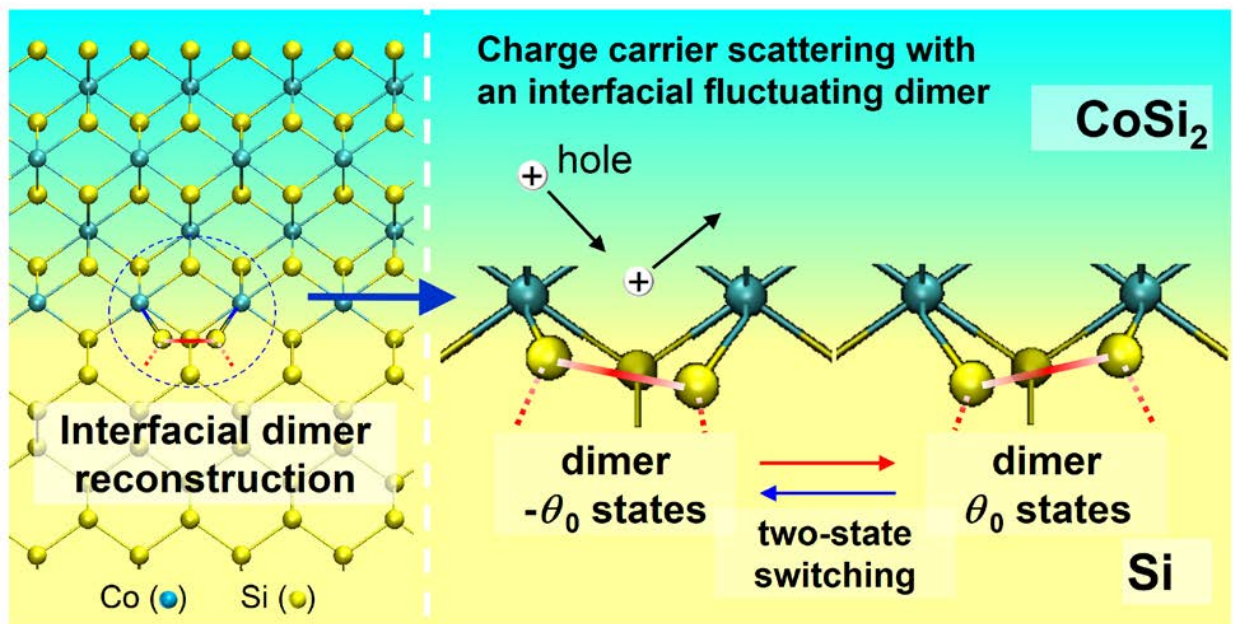

# Ultralow 1/$f$ Noise in a Heterostructure of Superconducting Epitaxial Cobalt-Disilicide Thin Film on Silicon

Shao-Pin Chiu,[†] Sheng-Shiuan Yeh,[†] Chien-Jyun Chiou,[‡] Yi-Chia Chou,*,[‡] Juhn-Jong Lin,*,[†,‡] and Chang-Chyi Tsuei[§]

[†]Institute of Physics, National Chiao Tung University, Hsinchu 300, Taiwan

[‡]Department of Electrophysics, National Chiao Tung University, Hsinchu 300, Taiwan

[§]IBM Thomas J. Watson Research Center Yorktown Heights, NY 10598, U.S.A.

*Email: ycchou@nctu.edu.tw

*Email: jjlin@mail.nctu.edu.tw

## S1. Superconducting transition and magnetoresistance for CoSi$_2$/Si(100) heterostructures

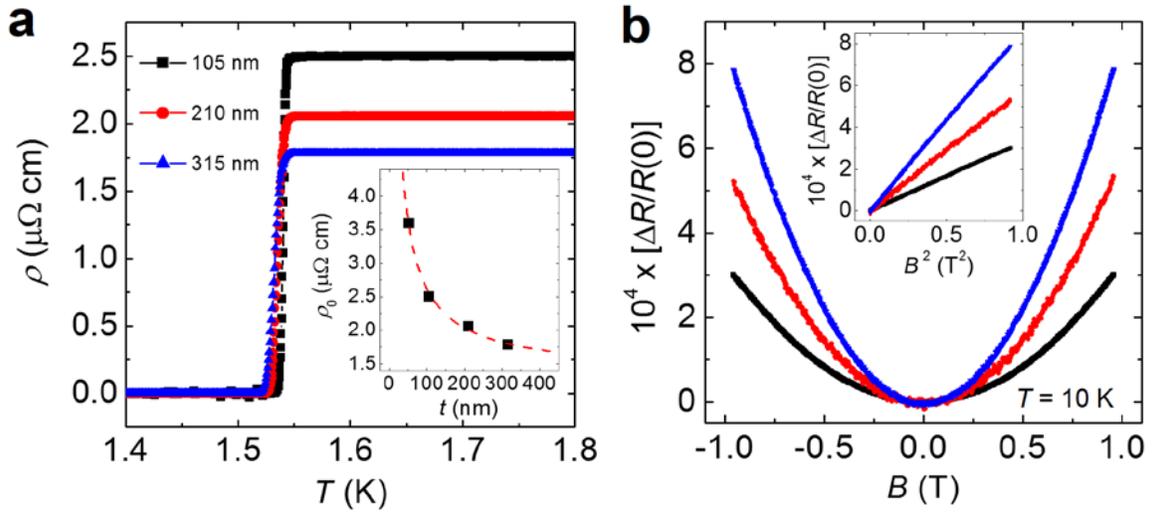

**Fig. S1 (a)** Resistivity as a function of temperature around superconducting transitions for three CoSi$_2$ films, as indicated. Inset: variation of residual resistivity with film thickness. **(b)** Normalized magnetoresistances for 105, 210, and 315 nm thick CoSi$_2$ films, as indicated. The data for the 105 nm, 90 min annealed CoSi$_2$ film (black curves) are also plotted in the Figs. 1(b) and 1(d) in the manuscript.

We have carried out additional electrical and magneto-transport measurements on 53, 210 and 315 nm thick CoSi$_2$ films on Si(100). Figure S1(a) shows very sharp superconducting transitions for three representative films, suggesting excellent sample quality. The inset shows the variation of residual resistivity $\rho_0$ with film thickness $t$. The decrease of $\rho_0$ with increasing $t$ implies that boundary scattering still plays a role even in our thickest film. By least-squares fitting to the Fuchs-Sondheimer theory [1,2] (the red dashed curve) as previously done by Hensel, *et al.* [3], we obtained the following parameters: bulk resistivity $\rho_{0,bulk} \simeq 1.48\ \mu\Omega\ cm$, bulk elastic mean free path $l_{e,bulk} \simeq 363\ nm$, and scattering specularity $p \simeq 0.45$. This $l_{e,bulk}$ value is close to the grain size in the lateral directions in our 105 nm $CoSi_2$ films, as mentioned in the manuscript.

Figure S1(b) shows that the magnitude of the normalized magnetoresistance (MR), $\Delta R / R(0) \approx (\mu B)^2$, increases with increasing film thicknesses, suggesting higher mobility $\mu$ in thicker films. Using the extracted $\mu$ values and following the calculation procedure described in the manuscript, we obtained $l_e = 121$, 142 and 175 nm in the 105, 210 and 315 nm thick films, respectively. This thickness dependent $l_e$ value implies and supports a relatively long $l_{e,bulk} \simeq 363\ nm$ mentioned above. Thus, for the 105 nm thick $CoSi_2$ films studied in this work, boundary scattering dominates the electrical transport. This result confirms that the measured $1/f$ noise originates from TLS fluctuators at interface.

## S2. Assignment of atomic positions and atomic model for the $CoSi_2/Si(100)$ interface

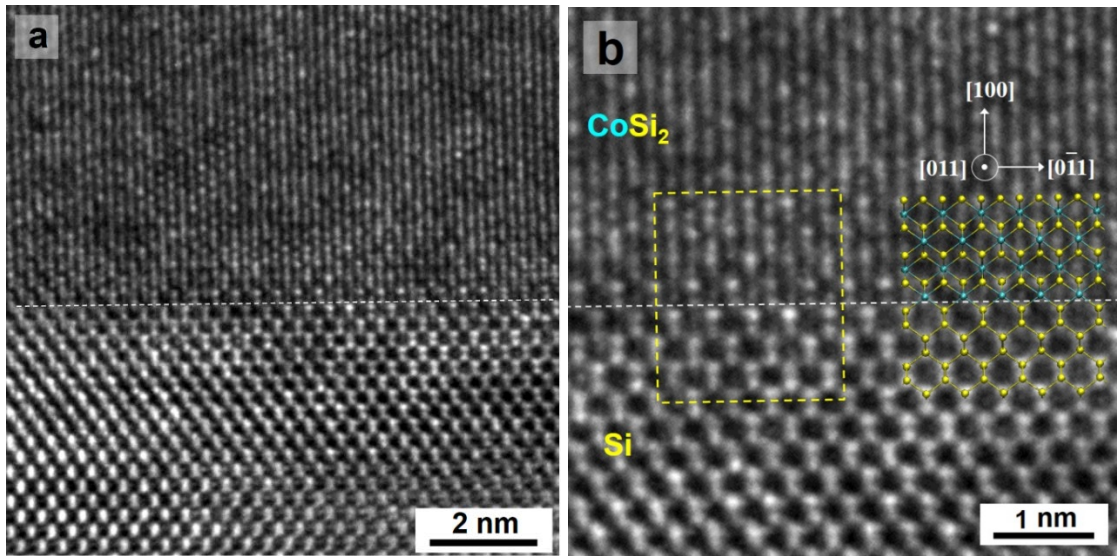

**Fig. S2** Cross-sectional HRTEM images for 90 min annealed $CoSi_2/Si$ sample. **(a)** The field size is $\approx 10\times 10\ nm^2$. The dashed line indicates the $CoSi_2$-Si boundary. **(b)** An enlarged image of (a). The field size is $\approx 5\times 5\ nm^2$.

We describe step by step how we identify the atomic positions at the $CoSi_2/Si$ interface from the HRTEM image. Figure S2(a) shows a cross-sectional HRTEM image with a $10\times 10\ nm^2$ field of view, which was taken with a magnification of 800 K, the resolution limit of our microscope (JEOL ARM200F). From the differences in contrast and the atomic arrangements between the upper and lower parts of this image, we identify the $CoSi_2$-Si interface by the dashed line, where the upper part being $CoSi_2$, and the lower part being Si. A partial area of $\approx 5\times5$ nm$^2$ field size was then cropped from Fig. S2(a) and enlarged in Fig. S2(b), with the boundary line remaining unchanged. The lower part of Fig. S2(b) shows a clear lattice image of the Si(011) plane, and thus allowing us to unambiguously identify the Si atomic positions. Moreover, from the diffraction pattern (DP) of the $CoSi_2$ part (the DP is shown in the inset of Fig.2(c) in the manuscript), we can clearly identify the orientation of the $CoSi_2$ lattice and, furthermore, construct a jointed atomic model with the Si lattice. Basing on the Si atomic positions and the boundary line, we map our atomic model on the HRTEM image in Fig. S2(b). The area enclosed by the yellow dashed square was cropped and enlarged again in Fig. S3(a). (This Fig. S3(a) is the same figure shown in the left panel of Fig.5(a) in the manuscript.) Finally, we map the same ball-and-stick model in Fig. S2(b) onto the TEM image of Fig. S3(a) to obtain Fig. S3(b).

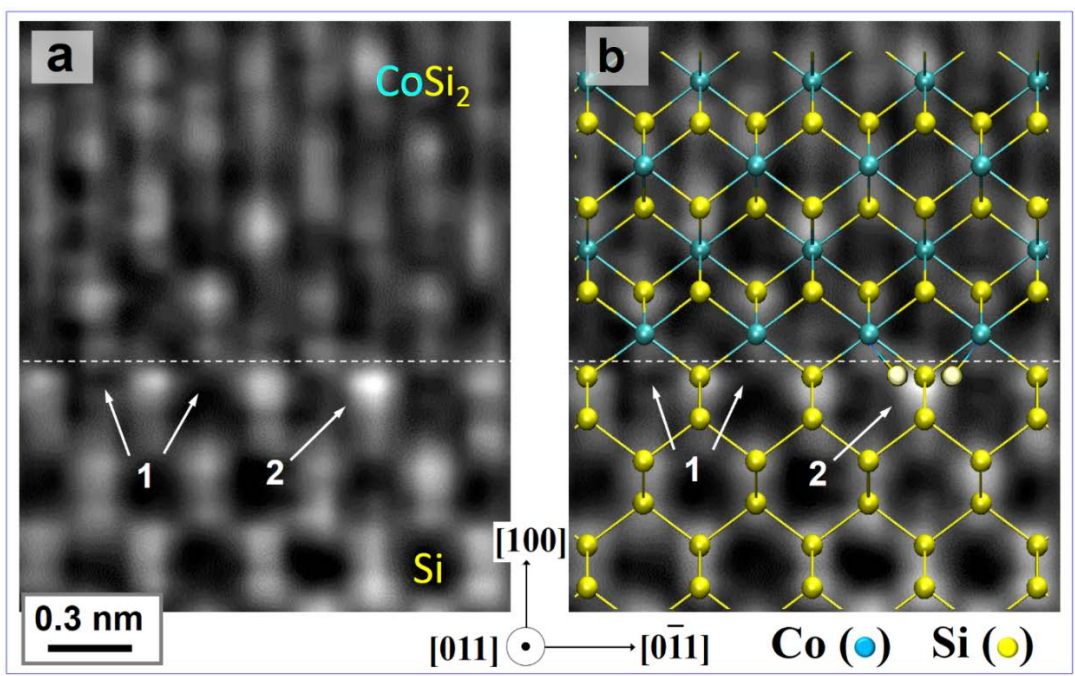

**Fig. S3** This figure is the same as the Fig. 5(a) in the manuscript. **(a)** A cross-sectional HRTEM image with epitaxy of $CoSi_2(100)$ on Si(100) taken from the yellow dashed box in Fig. S2(b). The field size is $\approx 1.6\times2$ nm$^2$. **(b)** A ball-and-stick model projected on the HRTEM image of (a).

In Fig. S3(b), the way for identifying the atomic positions at the interface is the same as above. The resolution for the $CoSi_2$ image is not ideal, probably owing to the oxidation during the transfer of sample, but it is still satisfactorily acceptable after enlargement for several times. In brief, our ball-and-stick model does well fit the $CoSi_2$/Si interface and the Si atomic positions. Hence, the brightest spots nearby the boundary dashed line can be modeled as excess Si atoms, leading to dimer reconstructions at the $CoSi_2$/Si interface. It should be noted that similar features in the TEM image with a corresponding dimer model explanation was first found and addressed in Ref. [4].